\documentclass[aps,pre,twocolumn,float]{revtex4}
\usepackage{amsmath,bm,epsfig}
\newcommand{\B}[1]{{\bm{#1}}}
\begin{document}

\title{Emergent Inter-particle Interactions in Thermal Amorphous Solids}
\author{Oleg Gendelman$^1$, Edan Lerner$^2$, Yoav G. Pollack$^3$,
Itamar Procaccia$^3$, Corrado Rainone$^3$ and Birte Riechers$^{3,4}$}
\affiliation{$^1$ Dept. of Mechanical Engineering, The Technion, Haifa, Israel,\\ $^2$Dept of Physics, University of Amsterdam, The Netherlands\\ $^3$Dept of Chemical Physics, The weizmann Institute of Science, Rehovot 76100, Israel\\
$^4$ permanent address: Physik. Inst. University of G\"ottingen, Germany}
\begin{abstract}
Amorphous media at finite temperatures, be them liquids, colloids or glasses, are made of interacting particles that
move chaotically due to thermal energy, colliding and scattering continuously off each other. When the average configuration in these systems relaxes only at long times, one can introduce {\em effective interactions} that keep the {\em mean positions} in mechanical equilibrium. We introduce a new framework to determine the effective force-laws that define an effective Hessian that can be employed to discuss stability properties and density of states of the amorphous system. We exemplify the approach with a thermal glass of hard spheres; these feel zero forces when not in contact and infinite forces when they touch. Close to jamming we recapture the effective interactions that at temperature $T$ depend on the gap $h$ between spheres as $T/h$  [C. Brito and M. Wyart, Europhys. Lett. 76 149 (2006)]. For hard spheres at lower densities or for systems whose binary bare interactions are longer ranged (at any density) the emergent force laws include ternary, quaternary and generally higher order many-body terms, leading to a temperature dependent effective Hessian.
\end{abstract}
\maketitle

{\bf Introduction}: The experimental determination of the inter-particle forces in amorphous glassy system
is a non-trivial challenge that has attracted a lot of effort, both in athermal granular systems \cite{05MB,14HMRA,06ZLWD}
and in thermal systems like colloids \cite{02BWQSPP,14LMCLIC,13RPW,13ZNSB,14ZRSTB}.
The aim of this communication is to introduce a new method to determine the emergent force laws of the ``effective interactions" between particles in thermal amorphous systems, especially in systems where direct measurements are either very difficult or even impossible. The suggested approach is only relevant for thermal amorphous systems that are ``arrested" in the sense that the thermal dynamics of every particle is restricted to vibrations within a cage. We thus aim at glassy systems, at temperatures below the glass transition, or colloids at densities sufficiently high to suppress diffusion for sufficiently long times. Thus the basic prerequisite to the discussion below is a possibility to measure
the {\em average positions} of the particles in the amorphous system, usually by averaging over the trajectory of each particle for times that are sufficiently long to produce a converged answer, but shorter than any relaxation time that destroys the cage structure to allow diffusion of the particles outside their cages. Denote then, in a system of $N$ particles at temperature $T$ the {\em average} positions of the particles as $\{\B r_i\}_{i=1}^N$. These average positions define a configuration that is time-independent. It is therefore legitimate and useful to ask what are the effective forces that are holding the configuration in force balance \cite{15DEM}. In the present Letter we exemplify the approach with thermal hard spheres near jamming and at lower densities. We will find that near jamming there exist almost only binary interactions and therefore binary effective forces $\B f_{ij}$ are sufficient \cite{16AFP}. For hard sphere at lower densities and for systems with binary longer ranged bare interactions,
emergent binary forces are {\em not} sufficient. Generically the emergent effective forces will include also ternary, quaternary
and generally higher order interactions.

Needless to say, the effective forces will generically turn out to be dependent on both the temperature and the density of the amorphous glass. But once determined, they can be manipulated in much the same way as the given forces of athermal systems, including the availability of an effective Hessian from which one can determine stability properties and density of states. The definition of such effective interactions opens a useful path for the discussion of thermal amorphous systems using the host of methods developed in the context of athermal systems.

{\bf Determining the effective forces, theory}: When only binary effective forces are needed, the necessary algorithm for determining them had been already developed elsewhere \cite{16GPP}, and it will only be summarized here briefly. Below we will extend the formalism
for cases that call for higher order effective interactions. For keeping the notation as simple as possible, we describe the algorithm for systems in 2-dimensions \cite{11MEMK}, with an obvious generalization to 3-dimensions.  The `mechanical equilibrium' constraints for the average positions read:
\begin{equation}
\B M|f_{ij}\rangle=0 \ ,
\label{MF}
\end{equation}
where $|f_{ij}\rangle$ is a vector of the {\em magnitudes} of the $c$ inter-particle central forces $\B f_{ij}$. For
simplicity we assume here periodic boundary conditions, otherwise walls introduce external forces that can be
easily taken into account, cf. Ref.~\cite{16GPP}. The matrix $\B M$ contains the $x$ and $y$ components of the
unit vectors $\B {\hat n_{ij}} \equiv \B r_{ij}/r_{ij}$ with $\B r_{ij}$ being the vector distances between the
particle positions $\B r_{ij} \equiv \B r_i - \B r_j$. Explicit examples of such a matrix can be found in
Ref.~\cite{15GPPSZ}.

Next the inter-particle force magnitudes are presented as Taylor-Laurent polynomials:
\begin{equation}
f^{AB}_{ij}=\overset{\ell_2}{\sum _{k=\ell_1}}a_k^{AB}\left(r_{ij}-r_0^{AB}\right)^k
\label{Taylor}
\end{equation}
where $\ell_1$, $\ell_2$ are the most negative and most positive powers in the expansion respectively. Below we will
denote the number of terms in the expansion as $\ell\equiv \ell_2-\ell_1+1$. $AB$ denotes the interaction type. For example in the case of a binary system these will be  (AA,BB,AB), as  determined by the nature of the particles $i,j$. In general we can have $n$ types of particles. $r_0^{AB}$ are the positions of the possible singularities around which we expand the forces for each type of interaction.
The coefficients $a_k^{AB}$ can be grouped into a vector $|a\rangle$ of size $n(n+1)\ell/2 $ and the force vector can now be written as:
\begin{equation}
|f_{ij}\rangle = \B S |a\rangle \\
\end{equation}
where $\B S$ is the appropriate $c\times n(n+1)\ell/2$ matrix containing the Tyalor-Laurent monomials.

To have a unique solution for the coefficients of the Taylor-Laurent expansion we need to fix one
scale parameter. While in experiment we would measure the pressure (and see Ref.~\cite{16GPP} for details),
in simulation of hard spheres we have to calculate the impulse applied in collisions
(divided by duration) $\sum\Delta \B p_{ij}/\Delta t$ where the sum is over all the collisions
of the same pair of particles, and $\Delta t$ is the time of measurement. Given the vector distances between the average position of the particles, say $\B r_{ij}$
we can calculate the virial $\upsilon$
as
\begin{equation}
\upsilon \equiv \langle\B r_{ij}| \sum\Delta \B p_{ij}/\Delta t\rangle \ .
\end{equation}
The problem of finding the effective forces then takes the form
\begin{equation}
\left(
\begin{array}{cc}
\B M \\
\langle r_{\text{ij}}| \\
\end{array}
\right)
\begin{array}{c}
\left|f_{\text{ij}}\rangle\right. \\
\end{array}
=
\left(
\begin{array}{c}
\B M \B S \\
\langle r_{\text{ij}}| \B S \\
\end{array}
\right)
\begin{array}{c}
|a\rangle \\
\end{array}
\equiv \B Y|a\rangle
=\left(
\begin{array}{c}
0 \\
: \\
\upsilon \\
\end{array}
\right)
\equiv |t\rangle \ ,
\end{equation}
where $\B Y$ is a $(2N+1) \times \ell$ matrix. We now multiply by $\B Y^T$ from the left
\begin{equation}
\B Y^T \B Y|a\rangle
=\B Y^T  |t\rangle
\label{ls}
\end{equation}
The set of unknown coefficient $|a\rangle$ is then solved for using standard Least-Squares methods.

One should note at this point that indeed the calculation of the impulse $\sum\Delta \B p_{ij}/\Delta t$ provides
a direct measurement of the effective forces between particles, see for example Refs.~\cite{06BW,09BW}. The procedure
proposed here provides however the emergent effective {\em force-laws}, and not just the forces. Moreover, while in numerical
simulations the measurement of the impulse is possible, it would be a hopeless proposition in any experimental
setting, where nevertheless the measurement of the pressure and mean positions are readily available. The method proposed here remains valid when the pressure instead of the virial is provided, cf. Ref.~\cite{16GPP}.

{\bf Example: thermal hard spheres:}
To exemplify the above procedure we chose a 2D hard-sphere event-driven simulation of over $10^7$ collisions with a
system size of $N = 400$ particles with periodic boundary conditions. The particles follow ballistic trajectories until they make contact and undergo an elastic collision. The particle radii $R$ are slightly poly-dispersed around a binary distribution with mean values and standard deviation of:
$\langle R_A\rangle = 0.5$, $\sigma_{R_A} = 0.0081$, $\langle R_B\rangle = 0.7$ and
$\sigma_{R_B} = 0.0123$.
A jammed state is used as the initial configuration to be expanded as explained below.
The system has constant volume $V$  and a constant energy $T=1$ (in units for which Boltzmann constant is unity), with the initial momenta $|p\rangle $ of the particles  chosen randomly (with constant distributions on $p_x$ and $p_y$ separately in the
interval $[-1,1)$, subject to the constraint that $T = m \langle p|p\rangle$ . In hard spheres the temperature only sets the time scales, so we took $T=1$ and ran the system at different densities as determined by the volume.
To analyze systems with different packing fractions, the jammed state is expanded during the initialization of the simulation. Expansions from an initial box of length $L_{init}$ to length $L$ were applied such that 
\begin{equation}
L = L_{init}\times(1+\epsilon) \ ,
\label{eps}
\end{equation}
with $\epsilon$ between $10^{-5}$ and $5.5\cdot10^{-2}$. The average positions were determined using averaging times that are well below the time for which particle diffusion  destroys the meaning of the mean positions, but higher than the typical time between collisions. In practice this means that we are limited in choosing the values of $\epsilon$ in Eq.~(\ref{eps}). For value of $\epsilon\geq 5.5\times 10^{-2}$ we could not determine the mean positions of the particles with
sufficient accuracy.
To ensure that early dynamics due to the initial expansion do not affect the analysis, the first configuration of a simulation is discarded.
To ensure that the ``noisy transition events" from one stable configuration to another \cite{06BW,09BW} do not affect the analysis, the first $10^4$ collisions in the analyzed stable configuration are discarded.

For densities very close to jamming one expects the effective forces to remain binary, \cite{06BW,09BW}. In hard spheres the energy is simply $T$. If one assumes that the only important scale is the gap $h$ between close-by particles, dimensional considerations predict that the effective forces would be simply $T/h$. As shown below,
this simple assumption is likely to fail at lower densities \cite{06BW,09BW}.
We first examine the efficacy of our proposed method in supporting this binary effective force-law.

{\bf Results for binary effective forces:}
Having computed the average positions of the centers of mass $\B r_i$ of all the $N$ particles, we
determined the vector distances $\B r_{ij}$ and relative gaps between particles as $h_{ij}$ where
\begin{equation}
h_{ij} \equiv r_{ij} - R_i-R_j \ .
\label{defh}
\end{equation}
Identifying in Eq.~(\ref{Taylor}) $r_0^{AB} = R_i^A+R_j^B$ we rewrite that equation in the form
\begin{equation}
f^{AB}_{ij}=\overset{\ell_2}{\sum _{k=\ell_1}}a_k^{AB}h_{ij}^k \ .
\label{Taylor2}
\end{equation}
This simplifies the $\B S$ matrix in the present problem to the monomials in the gap values $h_{ij}^k$.
For hard disks we expect that the effective interactions will depend only on the gap $h_{ij}$ and we can simplify things further by dropping the distinction between particles of different size and the superscripts $A$, $B$ and $AB$. Taking then (for example) six monomials in Eq.~\ref{Taylor2} with $\ell_1=-3$
and $\ell_2=+3$ (without a constant term!), we solve the problem set by Eq.~(\ref{ls}) and find that
$a_{-1} \approx T=1$ and all the other coefficients vanish to high accuracy (better than $10^{-6}$).
\begin{figure}
\hskip -1 cm
\includegraphics[width=0.45\textwidth]{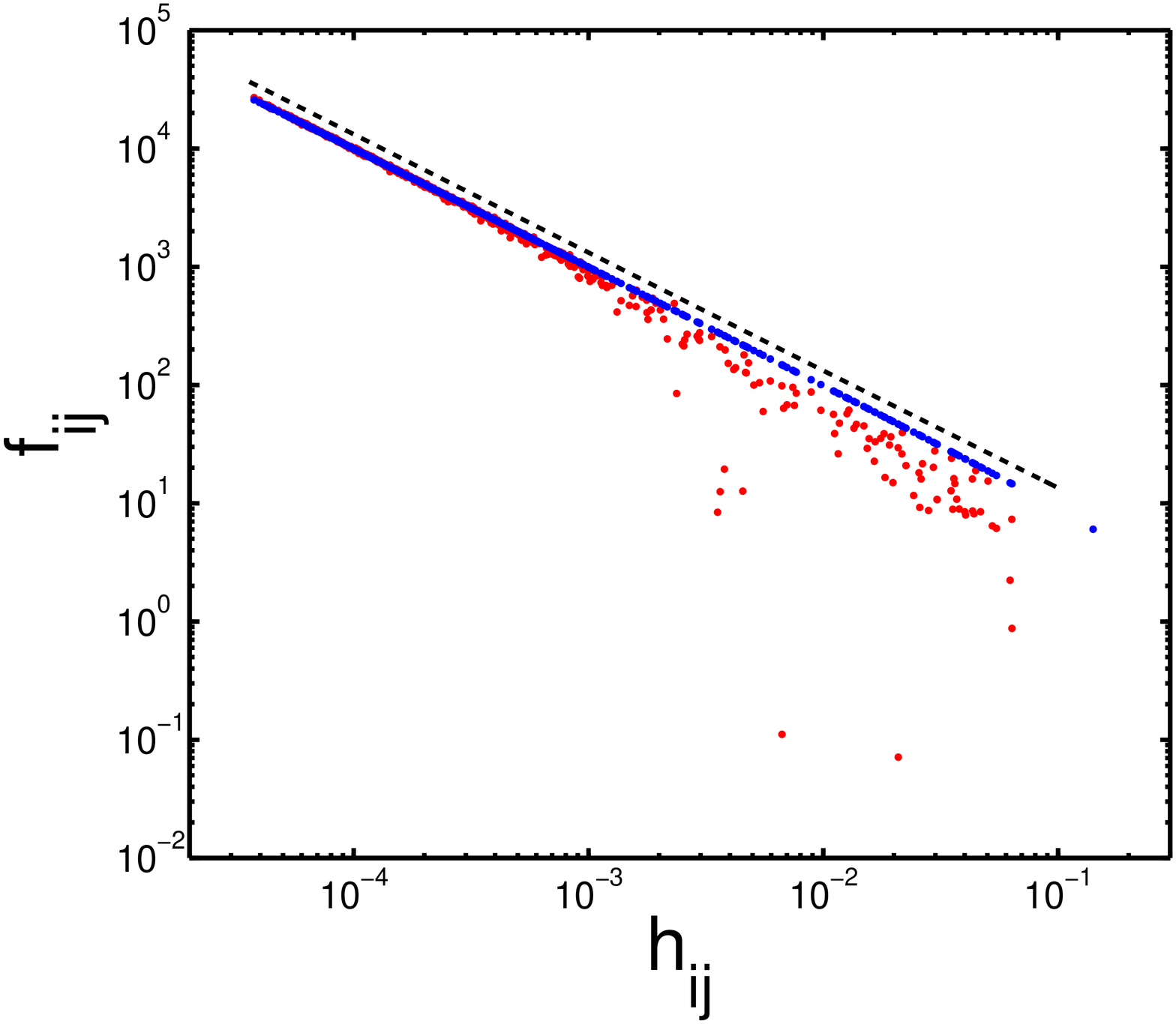}
\includegraphics[width=0.45\textwidth]{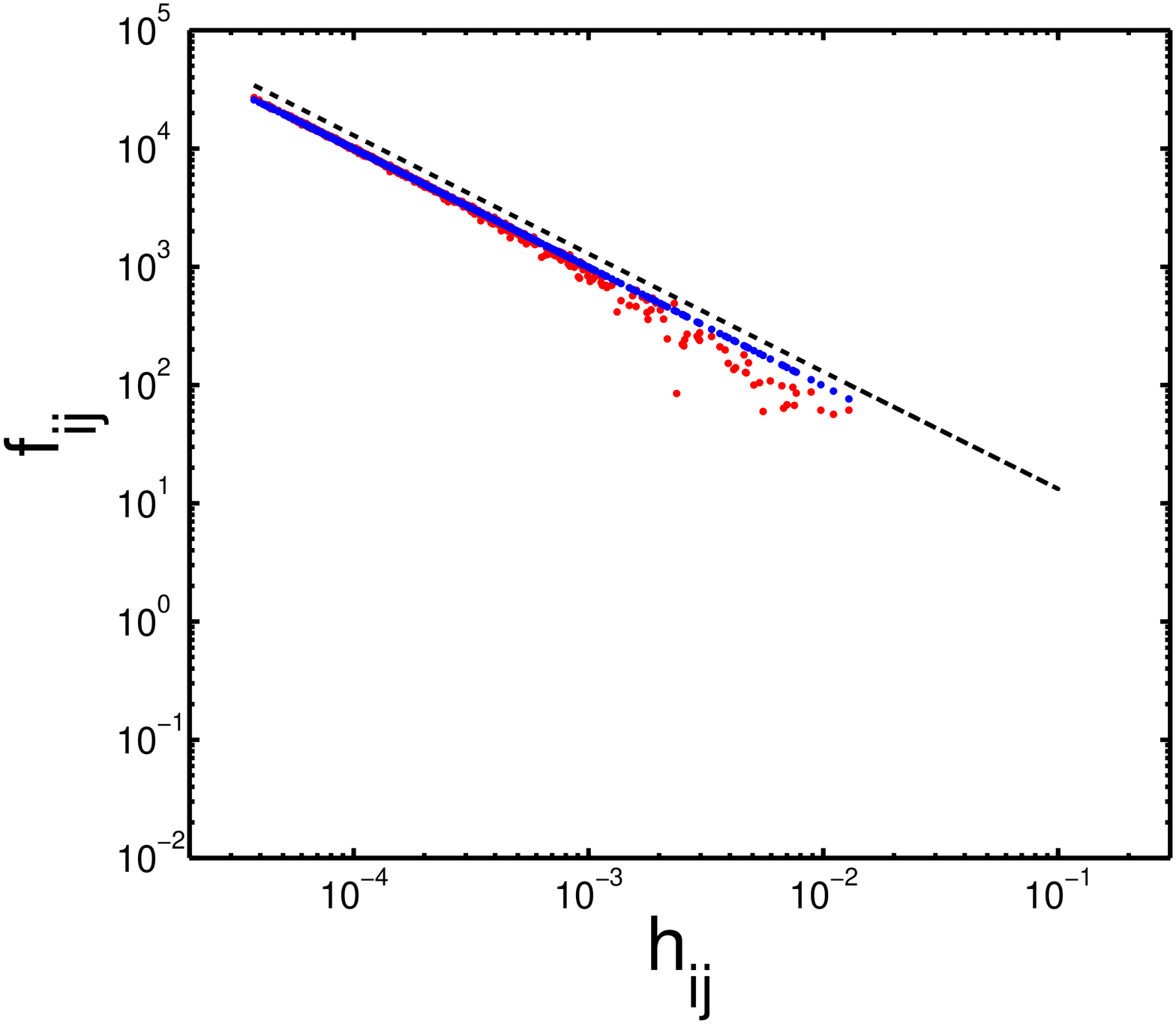}
\caption{Upper panel: Comparison of the computed effective forces for $\epsilon=10^{-4}$: in blue line we present the results obtained from the
present algorithm, i.e. $T/h_{ij}$. In red dots we show the results of estimating the forces from the direct momentum transfer
method. For small values of $h_{ij}$ the agreement is excellent, and it deteriorates at higher values. The
black dashed line is the expected result $C/h_{ij}$ with $C>T$ to allow comparison. Lower panel: the same data plotted after
excluding any particle pair which collided less than 100 times during the measurement
period and the corresponding measured forces.}
\label{forces}
\end{figure}
This result should be compared with the direct measurement of the effective forces between the particles, which,
as said above, can be evaluated from the momentum transfer, following verbatim the approach of Ref.~\cite{06BW,09BW}.
The blue line in the upper panel of Fig.~\ref{forces} represents the solution described here for $\epsilon=10^{-4}$, and the red dots are the estimates from the direct momentum transfer method. The black dashed line represents the law $C/h_{ij}$ and it coincides with the predicted emergent force law when $C=T$. The deviation of the red dots from the observed law indicate inaccuracies in the direct simulation that occur at larger values of $h_{ij}$.  We have checked and determined that the infrequent collisions between spheres separated by high gaps cause the decline in accuracy of the direct
measurements of the effective forces. The mean number of collisions between pairs in this simulation is 11547. By demanding that there should be at least 100 collisions
between an $ij$ pair whose force $f_{ij}$ is taken into account we obtain the comparison shown in the
lower panel of Fig.~\ref{forces}. The improvement in agreement is obvious.

The first sign that these results are not necessarily generic appeared while trying to repeat the same calculation for thermal glasses with Lennard-Jones bare forces. The effective binary forces contained a large number of Taylor-Laurent coefficients but failed to satisfy the mechanical constraints Eq.~(\ref{MF}).
The reason for this failure is deep. In the algorithm
proposed above we allowed only binary effective forces, with flexible Taylor-Laurent expansions, but only binary. This is appropriate in the dense hard spheres example since there are almost only binary interactions. In Lennard-Jones glasses even in very dense packings, any momentum exchange between two particles is strongly effected by other particles residing within the interaction distance. Accordingly,  we expect that
the emergent effective interactions will contain ternary, quaternary and higher order terms, depending on the density, the range of interactions and probably on the temperature. In fact, the same issues appear also
for hard spheres when we reduce the density, as we demonstrate next.

\begin{figure}
\hskip -1.5 cm
\includegraphics[width=0.55\textwidth]{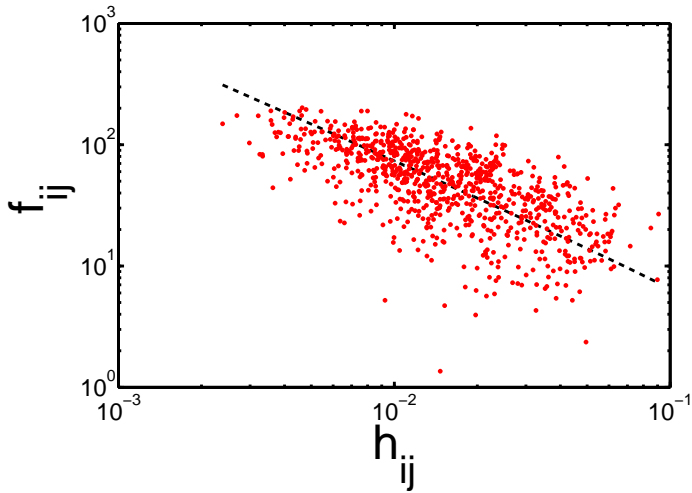}
\caption{Comparison of the measured forces (red dots) and the best fit to binary forces (dashed black line). The measured forces are no longer a graph of $h_{ij}$ since they
reflect the existence of multiple body interactions. The dashed line is a graph by construction, but obviously it does not represent the data well, and see Fig.~\ref{Fi}}
\label{direct}
\end{figure}

\begin{figure}
\hskip -1.5 cm
\includegraphics[width=0.40\textwidth]{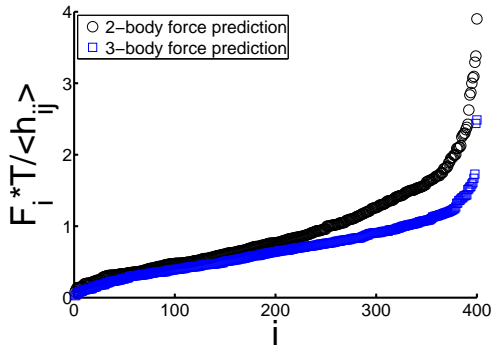}
\caption{The net forces on each, rescaled by $T/\langle h_{ij} \rangle$, plotted in increasing magnitude. In black circles we show the net forces computed from the best binary approximation. Blue squares represent the net forces computed with binary and some ternary contributions. The addition of even a limited number of ternary interactions improves the approximation.}
\label{Fi}
\end{figure}

{\bf Results for binary and ternary effective forces:} Reducing the density of the thermal hard spheres by using an expansion $\epsilon=10^{-2}$ in
Eq.~\ref{eps} changes the situation altogether. First, the direct measurements of the
forces does not yield a function of $h_{ij}$. This is demonstrated in Fig.~\ref{direct} where the red dots represent the forces measured directly from momentum transfer, plotted against
$h_{ij}$. The data scatter since the forces are not only functions of $h_{ij}$, they are
functions of more variables and do not fall on a graph as a function of $h_{ij}$.
Trying to fit the ``best" binary forces $f_{ij}$ results in the black dashed line in
Fig.~\ref{direct}. Obviously this resulting function does not do justice to the
scattered red dots; A good way to demonstrate the failure of the best binary approximation
is to compute the net force on each particle, $f_i=\sum_j f_{ij}$. This should vanish for every $i$ if the approximation were good. In Fig.~\ref{Fi} we show the net forces $f_i$ in order of increasing magnitude in black circles. Obviously the situation calls for the introduction of additional terms to the emergent effective forces.

For the sake of brevity we will demonstrate here how the addition of a limited type of 3-body terms leads to an improvement in satisfying the conditions of mechanical equilibrium of the mean positions. We will add only two types of terms, respecting the dimensionality
of the binary forces, i.e.
\begin{eqnarray}
f^{(2)}(h_{ij},h_{ik})&\equiv &\frac{a^{(2)}T}{(h_{ij}h_{ik})^{1/2}}\ , \nonumber\\
f^{(3)}(h_{ij},h_{ik},h_{jk})&\equiv& \frac{a^{(3)}T}{(h_{ij}h_{ik}h_{jk})^{1/3}} \ ,
\label{types}
\end{eqnarray}
where $a^{(2)}$ and $a^{(3)}$ are dimensionless coefficients to be determined.
We recognize that in general other 3-body and higher order terms may be necessary, but for the purposes of this
Letter it will be enough to the determine our effective forces in the present approximation as
\begin{equation}
\B f_{ij}=-\hat{n}_{ij}\Big(f_{ij}(h_{ij})+\sum _k \left[f^{(2)}(h_{ij},h_{ik}) +f^{(3)}(h_{ij},h_{ik},h_{jk}) \right]\Big)
\label{force-law}
\end{equation}
where the sum over $k$ goes over $k \neq ij$. One should recognize that the resulting forces $\B f_{ij}$ are {\em not} a function of $h_{ij}$ as is required. The method described above can be easily extended to determine the best fit in this form
and the result for the net forces when these terms are included are shown as the blue squares in Fig.~\ref{Fi}. The improvement with respect to the binary approximation is obvious, although convergence certainly requires additional terms.

In conclusion, we demonstrated a new approach based on measuring the average positions of particles in thermal amorphous systems in which the structural relaxation is slow. This allows us to define and compute emergent effective {\em force-laws}
that hold the system stable. In general the emergent forces include ternary, quaternary and in general higher order terms. Since the average positions are time-independent, we can now study how the
Hessian of the effective interactions can be used to predict the stability, the mechanical responses and
the density of states of thermal systems in much the same way as is done in athermal systems. The actual
emergent theory, including a full consideration of the many body interactions,  is beyond the scope of this communication and  will be presented elsewhere. In particular the convergence properties of this theory, both as a function of distance
from jamming and as a function of the order of the many-body terms call for an exciting and novel theory of thermal glasses.

\acknowledgments

This work had been supported in part by the Minerva Foundation, Munich, Germany. BR acknowledges support by the Deutsche Forschungsgemeinschaft through FOR 1394. We thank Matthieu Wyart for
useful email discussions.

\end{document}